# Predicting User Actions in Software Processes

Michael Deynet

Clausthal University of Technology
Julius-Albert-Str. 4
D-38678 Clausthal-Zellerfeld
Germany
michael.deynet@tu-clausthal.de

**Abstract.** This paper describes an approach for user (e.g. SW architect) assisting in software processes. The approach observes the user's action and tries to predict his next step. For this we use approaches in the area of machine learning (sequence learning) and adopt these for the use in software processes.

Keywords: Software engineering, Software process description languages, Software processes, Machine learning, Sequence prediction

## 1 Introduction

Software processes can help us to develop and deliver (complex) software systems. Hereby, a Software Process is a set of activities in a project which is executed to produce the software system. A Software Process Description is an abstract representation of a set of software processes. A Software Process Description Language is a language to describe a Software Process Description. Software Process Descriptions differ from each other in many things, e.g. the level of detail, the domain which is supported, the paradigm they implement.

The use of Software Process Description Languages has many advantages: The clear defined syntax and semantics make a tool-based interpretation possible. In software projects this is actually used for controlling, planning, and coordination of software projects.

A lot of Software Process Description Languages have been developed. Some implement one paradigm, for example rule based languages (pre-/postconditions), net based languages (petri nets, state machines), or imperative languages (based on programming languages). Others implement multiple paradigms.
There are advantages and disadvantages inherent in every paradigm:

Rule based languages have loosely coupled steps, which are flexibly combinable. The advantage is that the user (e.g. the developer, SW architect) of the process can execute the step in his own way. The disadvantage is that tool support during process execution is not possible for the user of the process. Thus, there is a lower benefit for the user.

Imperative and net based languages implement the step order directly. The step order corresponds to a generally valid order. Hereby, a tool based process execution is possible. The problem is that the order of the steps defined in the process description does not reflect the user's working method.
Our goal is to define a Process Description Language which prevents these disadvantages so that it is a) flexibly executable for the user (e.g. engineer, architect) but can nevertheless b) support the user with a tool by suggesting the next steps in the actual context.
The paper is structured as follows: Section 1.1 describes the related work in the area of software process languages and sequence learning. Section 2 gives an overview of our software process description language. In section 3 our approach to assist the user during process execution is shown. Section 4 shows our evaluation. The last section is the conclusion.

### 1.1 Related Work

Good summaries of work in the area of software process languages are [1], [2].

This paper contains approaches in the field of sequence prediction which is a subgroup of machine learning. Hartmann and Schreiber describe different sequence prediction algorithms in their paper [3]. The sequence prediction technique which is presented in this paper is based on IPAM [4] and Jacobs/Blockeel [5].

An overview paper of this work can be found here: [6].

## 2 Overview of the process description language

In this section we present a short overview of the concepts of our process modeling language. This language contains only elements which are needed to assist the user during process enactment. The language does not contain other elements like phases, roles, etc. We have designed a process language based on pre-/postcondition. One key element in our language is a **step** which is an activity during process execution. For example there are existing steps like "Map requirement to Component" or "Specify Component".

Every step has pre- and postconditions, that describe if the user can start or stop the step execution. Furthermore, steps can contain a set of **contexts**. This can be an **execution context** which is a subset of the product model (e.g. all classes within component c are in one execution context) or a "parameter" which describes a certain situation. This can be for example the implementation language, the used case tool, framework, the project, and so on. The number of steps is fix at project execution time; the number of contexts can vary (e.g. if a new product is added to the product model, one or more execution contexts are created corresponding to the process description). A more detailed description of our language can be found here [6].

# 3 User assistance

For user assistance we have developed an approach to predict the next step the user wants to start. Our approach observes the last couple of started steps (and corresponding contexts) and tries to build a database where identified sequences are stored. Our work is based on the work of Davison/Hirsh [4], and Jacobs/Blockeel [5].

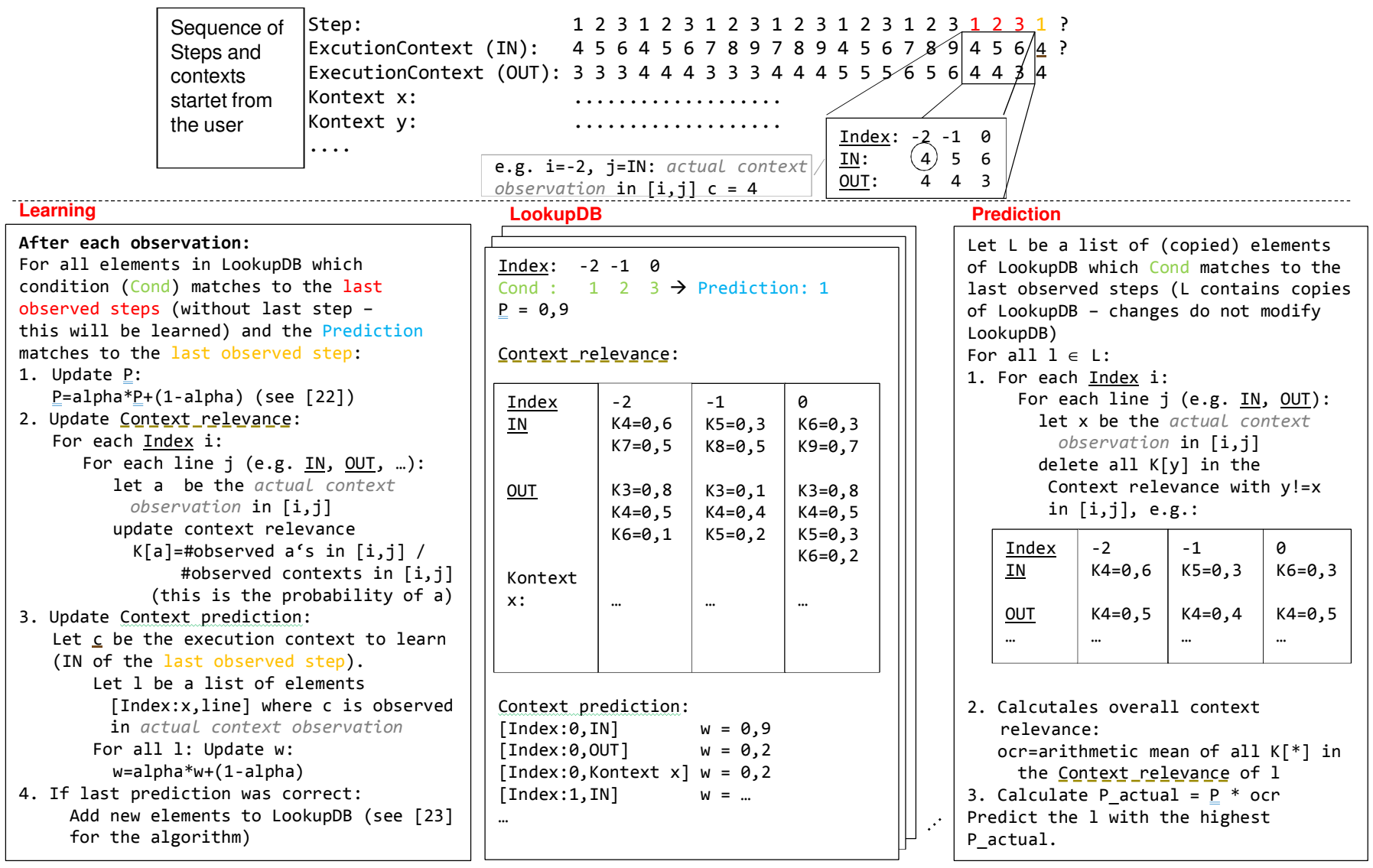

**Fig. 1.** Our approach for predicting the next user action

In the next subsection the results of the work of Davison/Hirsh and Jabobs/Blockked are presented.

## 3.1 Overview of the underlying work

IPAM is the work of Davison/Hirsh and addresses the prediction of UNIX commands. The approach implements a first order markov model, so the prediction which is made is based on the last observed element. IPAM stores unconditional and conditional probabilities in a database. After each observation the entries of the database are updated as follows: $P'(x|y) = \alpha \cdot P(x|y) + (1-\alpha)$ for $x$=current observation and $y$=last observation. The entries with $x \neq$ current observation are updated with: $P'(x|y) = \alpha \cdot P(x|y)$. $\alpha$ is a parameter between 0 and 1. Davison/Hirsh recommend a value of $\alpha \approx 0.8$.

The work of Jacobs/Blockeel is based on IPAM but implements a higher order markov model. If $\ldots y_1 y_0 x$ was observed and the prediction for $x$ was correct (e.g. $P(x|y_2 y_1 y_0)$ has highest probability) new entries are stored to the database. Let C be the set of suffixes of $\ldots y_1 y_0$ with $P(a|c) > 0$, for all $c \in C$. Let $l$ be longest Suffix of $\ldots y_1 y_0 x$ with $P(a|l) > 0$. The following new entries are stored to the database: $P(z|c \circ x) = P(z|l)$, for all $c \in C$ and $z \in$ observed elements.

### 3.2 Our approach

The central part of our approach is the LookupDB which stores the experience data figure 1 (middle box). Each of the elements of the LookupDB stores, among other things, a condition (*cond*), a step *prediction*, and a probability (*P*). This describes the probability (*P*) that the next step (*prediction*) will be started, given the occurrence of the step sequence *cond*. For example: An element of LookupDB can store the condition 1→2→3. When the steps 1→2→3 are observed, the probability that step 1 will be started is 0,9. Further, for each element in the condition sequence a set of relevant context probabilities are stored. Those are these contexts which were observed in the past.

For prediction (see figure 1 right) of the next step all relevant elements of LookupDB are taken and for each of these elements an "actual probability" is calculated. Here, our approach considers the learned context relevance in LookupDB and the actual observed contexts in the corresponding window.

After prediction and after observation our algorithm learns this observation by updating the conditional probabilities and context relevance of all matching elements in the LookupDB (see figure 1, left). In the following out approach is specified in a more detailed way.

First, let us define some relevant elements of our meta model (see [6]):
(1) Let $S$ be index set which represents the set of Steps
(2) Let $C$ be index set which represents the set of Contexts
(3) Let $CC$ be index set which represents the set of ContextClassifications
To address the past observed steps we need an index set I:
(4) Let $I$ be index set $[-n, \dots, 0]$.
The function *observation* returns the observed step at the index *i*, where *i=0* means the last observation, *i=-1* the last but one observation, et cetera (see fig. 1 top right):
(5) $observation \in I \rightarrow S$
Furthermore, we need a function which returns the observed context at index *i* and at the Context Classification *cc*:
(6) $observation2 \in (I \times CC) \rightarrow C$
In the example in fig. 1, *observation2(0,IN)* returns 6.

### 3.3 LookupDB

We define our LookupDB:
(7) Let *LOOKUPDB={0,1,2,…}* be Index Set
Each element of $LOOKUPDB$ represents an element of the LookupDB. Every element of our LookupDB has a condition *cond*. We define the function
(8) $cond \in (LOOKUPDB \times I) \rightarrow S$
that returns the conditional step of an element of the LookupDB at an index.
(9) $lenCond \in LOOKUPDB \rightarrow \mathbb{N}$
returns the length of the condition defined in (8). $cond(ldb, i)$ is defined for $i \in [-lenCond(ldb), 0]$.

The function prediction returns the predicted step of the element of the LookupDB:
(10) $prediction \in LOOKUPDB \to S$.
(11) $P \in LOOKUPDB \to [0,1]$
defines the probability of the element of the LookupDB that the predicted step occurs. The function *contextweight* returns a weight for each context classification (figure 1: the lines of the table, e.g. IN, OUT), index (figure 1: rows of the table) and context (number in the fields of the table) for each element of the LookupDB:
(12) $contextweight \in (LOOKUPDB \times CC \times I \times C) \to ([0,1],[0,1])$
$contextweight(ldb, cc, i, c) \coloneqq (x, y)\ with\ x =$ Number of observed $c$'s in $cc$ and $y =$ Number of all observed contexts ; (*x/y*) defines the probability that the context *c* at position *cc* and *i* is relevant)
Now, we define a function that returns 1 if a specified entry of LookupDB corresponds to the last observed steps:
(13) $match \in (LOOKUPDB \times \mathbb{N}) \to \{0,1\}$

$$match(ldb, offset) \coloneqq \begin{cases} 1 \text{ if } cond(ldb, q) = observation(q - offset) \text{ for all } q \in [-lenCond(lbd), 0] \\ 0 \text{ otherwise} \end{cases}$$

### 3.4 Prediction

For prediction we define the following functions:
(14) $getAMC \in \{(I \times CC \times C \times [0,1])\} \to [0,1]$
$getAMC(\{(i, cc, c, w)\}) \coloneqq amc$ with $amc :=$ arithmetic mean of all $w > \theta$ ($\theta$ is threshold) and
(15) $f \in LOOKUPTABLE \to \{(I \times CC \times C \times [0,1])\}$
$f(ldb) \coloneqq \{(i, cc, c, w)\}\ with\ i \in [-lenCond(ldb), 0]\ and\ c \coloneqq observation2(i, cc)$ for all $i$ and $cc$ and $w \coloneqq x/y\ with\ (x, y) \coloneqq contextweight(ldb, cc, i, c)$

The function *f* takes an entry of *LOOKUPDB* and returns a set of elements *(index,cc,c,w). w* corresponds to the contextweight of *ldb* for each index *i* and *cc* in *ldb* and for the corresponding *c* found in the observation.
The function *getAMC* takes a set of elements *(index,cc,c,w)* and returns the arithmetic mean of all *w*. *getAMC(f(ldb))* returns a “parameter” that the element of the LookupDB (and the learned context weights inside) corresponds to the observed contexts.

The function *getActualP* takes an entry of *LOOKUPDB* and calculates an actual P weight dependent on *ldb* and the actual observation (this weight is used to select the best entry of LOOKUPDB for prediction):
(16) $getActualP \in LOOKUPDB \to [0,1]$
$getActualP(ldb) \coloneqq getAMC\big(f(ldb)\big) * P(ldb)\}$
Now, we can describe the function to predict the next step:
(17) $makeprediction \in \{LOOKUPDB\} \to S$
$makeprediction(\{lbd\}) \coloneqq s)$ with
$s \coloneqq prediction(ldb)$ for maximal $getActualP(ldb)$

To predict the next step we call the function *makeprediction* with a set of all elements *ldb* of *LookupDB* with *match(ldb,0)*=1.

### 3.5 Learning

To update the LookupDB, the following steps are done (note: observation(0) is the step a prediction we made and we want to learn):

If there exists no $ldb \in LOOKUPDB$ with $(ldb, 1) \in lenCond$ and $((ldb, 0), observation(-1)) \in cond$ and $(ldb, observation(0)) \in prediction$ a new entry is added to the LookupDB (the entry with the shortest *cond*):

(18) $addEntry \in (\{LOOKUPDB\} \times cond \times lenCond \times prediction \times P) \rightarrow (\{LOOKUPDB\} \times cond \times lenCond \times prediction \times P)$

$addEntry(ldb_0, cond_0, lc_0, pre_0, p_0) \coloneqq (ldb_1, cond_1, lc_1, pre_1, p_1)$ with:

a) $n \notin ldb_0$ and $n \in ldb_1$
b) $((n, 0), observation(-1)) \notin cond_0$ and $((n, 0), observation(-1)) \in cond_1$
c) $(n, 1) \notin lc_0$ and $(n, 1) \in lc_1$
d) $(n, observation(0)) \notin pre_0$ and $(n, observation(0)) \in pre_1$
e) $(n, 1 - \alpha) \notin p_0$ and $(n, 1 - \alpha) \in p_1$

Let $LDB \subseteq LOOKUPDB$: $LDB = \{ldb | ldb \in LOOKUPDB, match(ldb, 1) = 1\}$
For each element $ldb \in LDB$ the following steps are done:

(19) $updateP \in (P \times LOOKUPDB) \rightarrow P$ with:

$updateP(p_0, ldb) \coloneqq p_1$ with $(ldb, pp) \in p_0$ and $\big(ldb, \alpha * pp + (1 - \alpha)\big) \in p_1$ (see [4])

(20) $updateContextProb \in (contextweight \times LOOKUPDB) \rightarrow contextweight$

$updateContextProb(cw_0, ldb) \coloneqq cw_1$ with $(ldb, cc, i, c, x, y) \in cw_0$ and $(ldb, cc, i, c, x + 1, y + 1) \in cw_1$ for all $c = observation2(i, cc)$ and $(ldb, cc, i, c, x + 1, y) \in cw_1$ for all $c \neq observation2(i, cc)$

This function updates the probabilities of the context information in *ldb*.

If the last prediction was correct new entries are added to the LookupDB according to the work of Jacobs et al. [5]. Let $Q \subseteq LOOKUPDB$ be a subset of *LOOKUPDB* with:

(21) $Q = \{ldb | ldb \in LOOKUPDB, cond(ldb, i) = observation(i, 1)$ for all $i \in [lenCond(ldb), 0]\}$

Let $L \subseteq LOOKUPDB$ be subset of *LOOKUPDB* with:

(22) $L = \{ldb | ldb \in LOOKUPDB, cond(ldb, i) = observation(i, 0)$ for all $i \in [lenCond(ldb), 0]\}$

Let *ll* be the element of *L* with the longest *lenCond* and *P(ll)*>0. The function *updateLOOKUPDB* is defined as:

(23) $updateLOOKUPDB \in (\{LOOKUPDB\} \times Q \times ll \times cond \times lenCond \times prediction \times P \times contextWeight) \rightarrow (\{LOOKUPDB\} \times cond \times lenCond \times prediction \times P \times contextWeight)$

$updateLOOKUPDB(LDB_0, q, l, c_0, lc_0, pre_0, p_0, cw_0) :=$ $(LDB_1, c_1, lc_1, pre_1, p_1, cw_1)$ with (see [5]):

a) $\{0,1,\dots,n\} \in LDB_0$ and $\{0,1,\dots,n,\dots m\} \in LDB_1$ with $|LDB_0| + |q| = |LDB_1|$

b) Let *ldiff* be $LDB_1 \backslash LDB_0$. For all $qq \in q$ there is an $ldiffel$ in $lediff$ with:
   a. $((q,i),s) \in c_0$ and $((ldiffel, i-1), s) \in c_1$ for all $i$ and $s$; and: $((ldiffel, 0), oberservation(0)) \in c_1$
   b. $(q,l) \in lc_0$ and $(ldiffel, l+1) \in lc_1$
   c. $(q,s) \in pre_0$ and $(ldiffel, s) \in pre_1$ for all $s$
   d. $(q,p) \in p_0$ $and$ $(ll, p2) \in p_0$ $and$ $(ldiffel, p2) \in p_1$

This function adds new entries to LookupDB by taking the entries which have predicted the observation correctly and "extending" the corresponding conditions by adding the observation (see [5] for detail).

## 4 Evaluation

For the evaluation of the approach described in section 3 we derived a realistic scenario. In this scenario the requirements of the system are existent. The goal is to develop an architecture (components, classes) and their implementation. The process description consists of 4 steps: 1) Identify component, 2) Map requirement to Component, 3) Specify component (refine requirement), 4) Implement component.

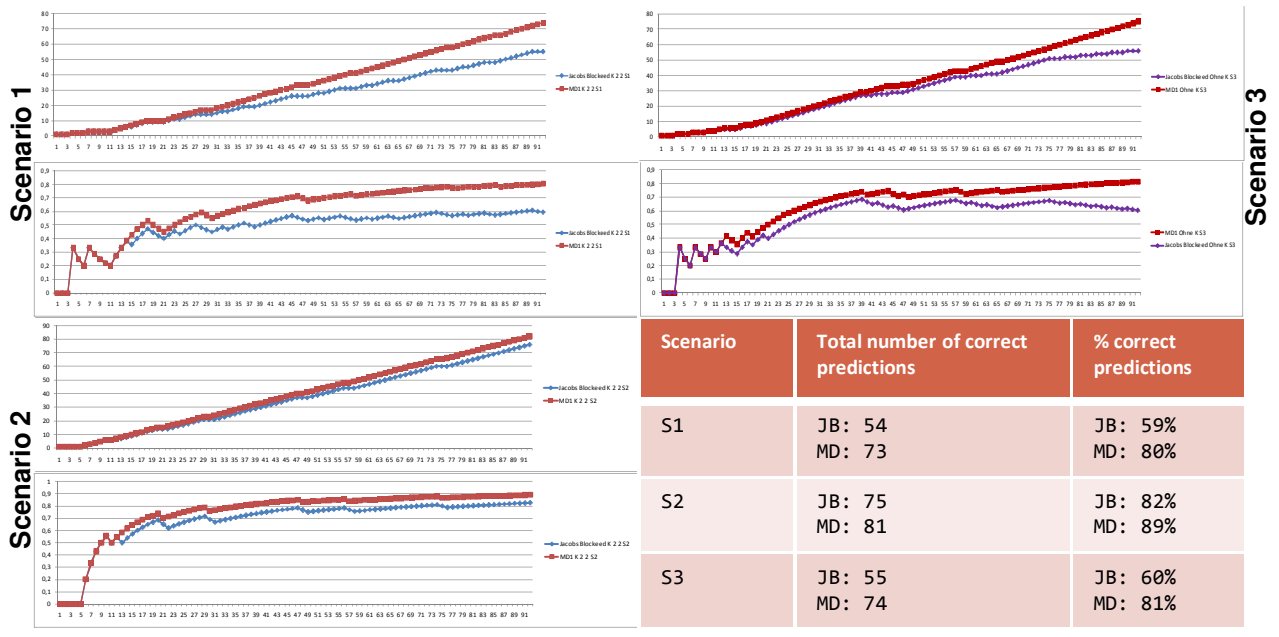

| Scenario | Total number of correct predictions | % correct predictions |
|---|---|---|
| S1 | JB: 54<br>MD: 73 | JB: 59%<br>MD: 80% |
| S2 | JB: 75<br>MD: 81 | JB: 82%<br>MD: 89% |
| S3 | JB: 55<br>MD: 74 | JB: 60%<br>MD: 81% |

**Fig. 2.** Evaluation results (red: Our approach; blue: Jacobs Blockeed)

In the system two types of components are existing: a) Complex/hardware related components. Here, the engineer has a prototypical method to develop the component (steps 2-4 are executed sequentially). b) Components which classes have a high coupling. Here, the engineer has a broad design method (step1; all steps 2; all steps 3; all steps 4).

For evaluation three sequences of steps (with corresponding contexts) were build: i) Development of components only of the type a, ii) Development of components only of the type b, and iii) random mix of a and b.

Our approach is compared with the algorithm from Jacobs/Blockeel [5]. The results are shown in figures 2. This figure contains two graphs for each scenario: The first (top) describes the total number of correct predictions and the second describes the percentage distribution of correct prediction. The Jacobs Blockeel approach is shown with blue color and our approach is shown with red color.

In all scenarios our approach predicts better than the Jacobs Blockeel algorithm. Remarkable is that in scenario 3 (the "real world" scenario) our approach predicts the steps substantially better than the JB approach. After 2/3 of all steps our algorithm predicts always correct. On the other hand the algorithm of JB "drifts" to 65% correct predictions.

## 5 Conclusions, Further Work

In this paper we described an approach to assist users by predicting the next step the user starts during process enactment. We evaluated this work by defining scenarios and we compared our approach with the core algorithm we have adopted. The results show that our approach predicts continuously better than the original algorithm.

We plan to evaluate our approach with "real project data" Furthermore, we'll integrate our concepts of the process language in a standard language/meta model and implement an integrated tool support for our prediction approach.

Other interesting extensions of our work could be: a) Supporting novice users (e.g. user works in a new project/new company) by providing the experience data of other users, b) Using the experience data (of the users of one or more projects) for organization-wide process improvement (e.g. derive a standard process description (standard sequence of steps; e.g. for a handbook) out of the experience data).

**Acknowledgments.** Thanks to the two reviewers for their detailed comments.

## 6 References

[1] Kamal Zuhairi Zamli, „Process Modeling Languages: A Literature Review", Dez-2001. [Online]. Available: http://myais.fsktm.um.edu.my/278/. [Accessed: 13-Jan-2009].

[2] S. T. Acuna, „Software Process Modelling". .

[3] M. Hartmann und D. Schreiber, „Prediction algorithms for user actions", *Proceedings of Lernen Wissen Adaption, ABIS*, S. 349–354, 2007.

[4] B. D. Davison und H. Hirsh, „Predicting sequences of user actions", in *Notes of the AAAI/ICML 1998 Workshop on Predicting the Future: AI Approaches to Time-Series Analysis*, 1998.

[5] N. Jacobs, H. Blockeel, A. Celestijnenlaan, und others, „Sequence prediction with mixed order Markov chains", in *In Proceedings of the Belgian/Dutch Conference on Artificial Intelligence*, 2002.

[6] M. Deynet, „User-Centric Process Descriptions", presented at the 3rd International Conference on Software Technology and Engineering ICSTE 2011, Malaysia.